\documentclass[useAMS,usenatbib,a4]{mn2e}

\usepackage{times}
\usepackage{graphicx,bm,amssymb}
\usepackage{ulem}
\usepackage{epsfig}
\usepackage{amssymb}
\usepackage{natbib}
\usepackage{pstricks}
\usepackage{psfrag}
\usepackage[colorlinks=true,linkcolor=blue,citecolor=blue]{hyperref}
\def\apj{ApJ}
\def\apjl{ApJ}
\def\apjs{ApJS}
\def\aap{A\&A}
\def\jcap{J. Cosmology Astropart. Phys.}
\def\mnras{MNRAS}
\def\prd{Phys.~Rev.~D}
\def\rmxaa{Rev. Mexicana Astron. Astrofis.}
\def\nat{Nature}
\def\physrep{Phys.~Rep.}

\newcommand{\be}{\begin{equation}}
\newcommand{\ee}{\end{equation}}
\newcommand{\bary}{\begin{eqnarray}}
\newcommand{\eary}{\end{eqnarray}}

\def\bi{\begin{itemize}}
\def\ei{\end{itemize}}
\def\lsim{\mathrel{\rlap{\lower3pt\hbox{\hskip1pt$\sim$}}
     \raise1pt\hbox{$<$}}} 
\def\gsim{\mathrel{\rlap{\lower3pt\hbox{\hskip1pt$\sim$}}
     \raise1pt\hbox{$>$}}} 

\begin{document}
\title[Hypercritical accretion phase and neutrino expectation in the evolution of Cassiopeia A]
{Hypercritical accretion  phase and neutrino expectation in the evolution of Cassiopeia A}
\author[N. Fraija and C. G. Bernal]
  {N.~Fraija,$^1$\thanks{E-mail:nifraija@astro.unam.mx. Luc Binette-Fundaci\'on UNAM Fellow.} and  C.~G. Bernal,$^2$\thanks{E-mail:astrogio@if.uff.br} \\
    $^1$Instituto de Astronom\' ia, Universidad Nacional Aut\'onoma de M\'exico, Circuito Exterior,
C.U., A. Postal 70-264, 04510 M\'exico D.F., M\'exico\\
        $^2$ Instituto de Fisica, Universidade Federal Fluminense, Av. Gal. Milton Tavares de Souza s/n, Gragoata, Niteroi, 24210-346, Brazil}
                
\maketitle
\date{\today} 	
\begin{abstract}
Cassiopeia A the youngest supernova remnant known in the Milky Way is one of the brightest radio sources in the sky and a unique laboratory for supernova physics.  Although its compact remnant was discovered in 1999 by the Chandra X-Ray Observatory, nowadays it is widely accepted that a neutron star lies in the center of this supernova remnant. In addition, new observations suggest that such neutron star with a low magnetic field and evidence of a carbon atmosphere could have suffered a hypercritical accretion phase seconds after the explosion. Considering this hypercritical accretion episode, we compute the neutrino cooling effect, the number of events and neutrino flavor ratios expected on Hyper-Kamiokande Experiment.  The neutrino cooling effect (the emissivity and luminosity of neutrinos) is obtained through numerical simulations performed in a customized version of the FLASH code.  Based on these simulations, we forecast that  the number of events expected on the Hyper-Kamiokande Experiment  is around 3195.  Similarly, we estimate the neutrino flavor ratios to be detected considering the neutrino effective potential due to the  thermal and magnetized plasma and thanks to the envelope of the star. It is worth noting that our estimates correspond to the only trustworthy method for verifying the hypercritical phase and although this episode took place 330 years ago, at present supernova remnants with these similarities might occur thus confirming our predictions for this phase.
%
%
\end{abstract}
%
%
%
\section{Introduction}
At a distance of $d_z=3.4^{+0.3}_{-0.1}$ kpc,  Cassiopeia A (Cas A) is one of the closest and youngest known supernova remnants  \citep{1995ApJ...440..706R,2004ApJ...605..830C}.  It was related to the supernova observed by John Flamsteed in 1680 \citep{1980JHA....11....1A}. As was reported by \cite{Krause2008}, Cas A supernova remnant was produced in a IIb type supernova
explosion. The progenitors of such explosions, as was showed by \cite{Chevalier2005}, are red supergiants which already have lost their hydrogen envelope via a powerful stellar wind. Then, it is inferred that such progenitors with masses of the order of $20\:\mathrm{M_{\odot}}$ must have a helium core of $6\:\mathrm{M_{\odot}}$ and an iron core of $1.45\:\mathrm{M_{\odot}}$. This scenario should produce a neutron star with $1.4\:\mathrm{M_{\odot}}$. It seems to be the case of Cas A.\\
Studying the atmosphere surrounding the neutron star of Cas A, \cite{2009Natur.462...71H} found evidence of carbon in the composition of this atmosphere. It suggests that a strong accretion could have occurred after the neutron star formation.  In the core-collapse framework other evidences related to the low strength of magnetic field, i) the non-detection of the spectral features related to the electron cyclotron resonance in the Chandra energy range (which is usually associated to a $B\simeq (1- 5)\times 10^{11}$ G), ii)   the absence of a visible pulsar wind nebula and iii) no samples of magnetospheric activities (such as radio or $\gamma$-ray fluxes) from the direction of Cas A, as found in pulsars with $B\geq 10^{12}$ G,  suggest that the strong magnetic field was buried because of a strong accretion \citep{1995ApJ...440L..77M}.   If the core-collapse paradigm is correct, the shock wave sweeps the outer layers of the progenitor until it encounters a discontinuity in density. At this point, a reverse shock is generated leading to a fallback episode which allows to deposit large amounts of material on the newborn neutron star surface. As a consequence of the fallback episode,  the magnetic field is submerged within the new crust formed during such episode \citep{Bernal2013}.\\
In the hypercritical phase,  super-Eddington accretion gives rise to the material of core-collapse to become dense enough so that photons are trapped. Therefore, the gravitational energy is carried out due to neutrino emission in the formation of the newborn neutron star.  When the accretion rate $\dot{M}$ exceeds by several orders of magnitude the Eddington accretion rate $\dot M_{Edd} \sim 10^{18}$ gr s$^{-1}$ so-called "Eddington limit", then it is denoted as hypercritical accretion \citep{Blondin1986}, and the accretion rate required is higher than $\sim 10^3 \, \dot{M}_{Edd}$.\\
In the hypercritical accretion episode (seconds after the supernova explosion), due to the inverse beta decay, electron-positron annihilation ($e^-+e^+\to Z\to \nu_j+\bar{\nu}_j$) and nucleon-nucleon bremsstrahlung ($N+N\to N+N+\nu_j+\bar{\nu}_j$) for $j=e,\nu,\tau$, thermal neutrinos (with energies in the range 5 MeV$ \leq E_\nu \leq$ 30 MeV) will be produced at the core and then oscillate in their propagation through the collapsing star.  The resonant conversion (from one flavor to another)  and the  properties of these neutrinos will get modified due to neutrino effective potential generated by the magnetized and thermal plasma \citep{2009JCAP...11..024S, 2009PhRvD..80c3009S}, and later by the collapsing and surrounding material of the progenitor.  It is worth noting that, although neutrino cannot couple directly to the magnetic field, its effect can be felt through the coupling to charge particles in the thermal plasma. Therefore, the values of neutrino  flavor ratios created at the core will be different from the values on the surface of the star and on Earth.\\
Recently,  \citet{2014MNRAS.442..239F} revisited Chevalier's model in a numerical framework to focus on the neutrino cooling effect on the  SN1987A fall-back dynamics. They estimated the number of neutrinos that must have been  seen on the Super-Kamiokande neutrino experiment,  hours after the $\sim$ 20 events detected from the direction of SN1987A. In the current work, we consider that Cas A supernova remnant underwent the hypercritical accretion episode in its early life as neutron star.  We present a numerical study of hypercritical accretion of matter onto the neutron star surface.  As an evidence of this episode we estimate the neutrino luminosity, the expected number of neutrinos and the neutrinos flavor ratio on  Earth.  We employ the FLASH code to carry out the numerical simulations of the reverse shock and the complex dynamics near the stellar surface, including neutrino cooling processes, a more detailed equation of state, and an additional degree of freedom in the system.  The paper is arranged as follows. In section \ref{section2} we describe the dynamics of the hypercritical accretion phase. In section \ref{section3} we study the thermal neutrino expectation. In section \ref{section4} we discuss the results and a brief conclusions is drawn in section \ref{section5}. 
\section{Dynamics of the hypercritical accretion phase} 
\label{section2}
In the framework of the stellar core-collapse, a reverse shock is generated when the expanding shock wave encounters dense outer layers on its path. Whereas the forward shock advances out of the star, the reverse shock leads to a fall-back phase which can induce the hypercritical accretion onto the newborn neutron star surface during the following 3 hours after the explosion. This scenario is only possible if the progenitor has a tenuous H/He envelope surrounding a dense He core, as suggested by \citet{2009Natur.462...71H}. In this approach, the neutrino emission determines the formation of a quasi-hydrostatic envelope around the newborn neutron star in the hypercritical phase. 
In the above mentioned scenario, we can identify four dynamically different regions: (i) a thin crust surrounding the newborn neutron star, which is formed during the hypercritical phase; (ii) an envelope in quasi-hydrostatic equilibrium formed by the third expansive shock; (iii) a free-fall region above this envelope; and (iv) the external layers of the progenitor.
Now, we are going to consider the dynamics inside each region around the compact remnant.
 \subsection{The new crust on the neutron star surface}
\noindent 

When the hypercritical phase is occurring, the piled material builds a small region at the same height scale where the neutrino cooling takes place.  It forms a new thin crust with a strong magnetic field immersed within it. In this crust formed at a region  $r_{ns}  \leq r \leq (r_{ns}+r_{c}$) on the  surface of the newborn neutron star surface, photon field radiation and neutrinos are thermalized to a value in the range  1 MeV $\leq T \leq$  5 MeV.  Here, $r_{ns}=10^{6}$ cm is the neutron star radius and $r_{c}\sim 200-400$ m is the size of the new thin crust. The energy density and neutrino emissivity are given by \citep{Dicus1972}
\be
U_B(T)=3.77\times10^{26}\left(\frac{T}{\mathrm{MeV}}\right)^4\:\mathrm{erg\, cm^{-3}},
\label{equation8}
\ee
and
\be
\dot{\epsilon}_{\nu}=0.97\times10^{25}\left(\frac{T}{\mathrm{MeV}}\right)^{9}\:\mathrm{erg\, s^{-1}\, cm^{-3}},
\label{equation9}
\ee 
respectively.  At a first approach and due to the high dependence of the pair annihilation process on pressure near the stellar surface, the neutrino emissivity can be written as  
\be
\dot{\epsilon}_{\nu} = 1.83 \times 10^{-34} p^{2.25}\: \mathrm{erg \, cm^{-3} \, s^{-1}}\,,
\label{equation6}
\ee
\noindent where the total pressure $p$ includes (besides the gas pressure) the pressure due to electron/positron pairs which is defined by
\be
p_{e^{-}e^{+}}=\frac{11}{4}\left(\frac{aT^{4}}{3}\right)\,,
\ee
with $a$ the radiation constant. It is worth noting that, in the hypercritical phase, the magnetic field may be submerged and confined into the new crust (formed by the piled material).   The strength of the magnetic field could reach values  in the range of $10^{11}\,{\rm G} \leq B \leq 10^{13}\,{\rm G}$. Also, the density and pressure of radial profiles can be described as power laws, whereas the mean velocity is null in this region.
\subsection{Quasi-hydrostatic envelope}
The reverse shock induces the hypercritical accretion onto the newborn neutron star surface through the fall-back episode. This accreted material will bounce against  the surface of the newly born neutron star, building a new expansive shock. This expansive shock builds an envelope in quasi-hydrostatic equilibrium, with free falling material raining over it. The structure of such envelope in quasi-hydrostatic equilibrium is given by \citep{Chevalier1989}
\be
\rho_{qhe}=\rho_{s}\left(\frac{r_{s}}{r}\right)^{3},\,\,\,\,\, p_{qhe}=p_{s}\left(\frac{r_{s}}{r}\right)^{4},\,\,\,\,\, v_{qhe}=v_{s}\left(\frac{r_{s}}{r}\right)^{-1},
\label{equation2}
\ee
\noindent where the subscript \textit{s} refers to the value of density ($\rho_{s}$), pressure ($p_{s}$) and velocity  ($v_{s}$) at the shock front. Given the typical values of the neutron star:  a mass $M=1.4\,\mathrm{M_{\odot}}$, an accretion rate $\dot{M}=500\,M_{\odot}\, yr^{-1}$ and a radius $r_{ns}=10^{6}$ cm, the shock radius is   
{\small
\bary\label{rs}
r_{s}&\simeq&7.73\times10^{8}\mathrm{cm} \left(\frac{M}{1.4\,\mathrm{M_{\odot}}}\right)^{-0.04}\left(\frac{r_{ns}}{10^{6}\mathrm{\, cm}}\right)^{1.48}\cr
&&\hspace{3.6cm}\times\left(\frac{\dot{M}}{\mathrm{500\,M_{\odot}\, yr^{-1}}}\right)^{-0.37}.
\eary
}
\noindent The pressure $p_{s}$ and density $\rho_{s}$ are determined by the strong shock condition $p_{s}=\frac{6}{7}\rho_{0}v_{0}^{2}$ and $\rho_{s} = 7 \rho_{0}$, respectively and $v_{s}$ by the mass conservation $v_{s} = -\frac{1}{7} v_{0}$.   Here $\rho_{0}$ and $v_{0}$ are the density and velocity just outside the shock, respectively.  In this case, the radial distance of the accretion shock that is controlled by the energy balance between the accretion power and the integrated neutrino losses is given by
\be
\frac{GM\dot{M}}{R} = \int_R^\infty \dot{\epsilon}_{\nu}(r) dr.
\label{equation4}
\ee
\noindent \citet{Chevalier1989} approached the size of the envelope as $R/4$ for the pressure profile in an atmosphere in quasi-hydrostatic equilibrium. With this, and using the analytical formula of \cite{Dicus1972} for the neutrino emissivity, the shock radius can be calculated through the energy balance as
\be
\frac{GM\dot{M}}{R}=4\pi R^{2}\left(\frac{R}{4}\right)\dot{\epsilon}_{\nu}.
\label{equation5}
\ee 
\noindent Then, the high pressure near the neutron star surface becomes 
\be
p_{ns}\simeq1.86\times 10^{-12} \mathrm {dyn\, cm^{-2}} \dot{M}\, r_{s}^{3/2}\,.
\ee
It allows the pair neutrino process to be the dominant mechanism in the neutrino cooling.  From the strong shock conditions and eq. (\ref{equation2}), we get 
\be\label{rho2}
\rho_{qhe}(r)= 7.7 \times 10^2 \,\left(\frac{r_{s}}{r}\right)^{3}~{\rm g~cm}^{-3}\,,
\ee
where the quasi-hydrostatic envelope radius lies in the range $(r_{ns}+r_{c}) \leq r \leq r_s$.
\subsection{Free fall zone}
\label{FF-zone}
In this case, material at $r_s\leq r \leq r_h$ begins to fall with the velocity and density profiles given by
\be
v_{ff}(r)=\sqrt{\frac{2GM}{r}},\qquad{\rm and}\qquad \rho_{ff}(r)=\frac{\dot{M}}{4\pi r^{2}v(r)}\,.
\label{equation1}
\ee
Taking the typical values for such object, $M = 1.4 ~{\rm M_\odot}$ and $\dot{M}= 500 ~{\rm M_{\odot}}\, {\rm yr^{-1}}$,  the velocity is written as
\be
v_{ff}(r)=7.81\times10^{7}\,\biggl(\frac{r}{r_h} \biggr)^{-1/2}\:\mathrm{cm\, s^{-1}}\,,
\label{equation13}
\ee 
and the density of material in free fall as
\be\label{rho3}
\rho_{ff}(r)=5.74\times 10^{-2}\,\biggl(\frac{r}{r_h} \biggr)^{-3/2}~{\rm g~cm}^{-3}\,,
\ee
where $r_h= 6.3\times 10^{10}$ cm.
\subsection{External layers}
By considering a supergiant profile (typical for a pre-supernova),  the density of the outer layers can be described  as  \citep{che89}
\begin{eqnarray}
\rho_{el}(r)&=&3.4\times10^{-5}\mathrm{g\,cm^{-3}}\cr
&&\hspace{0.9cm}\times\cases{
\left(\frac{R_\star}{r}\right)^{n_1}; & $r_h< r < r_b$,\cr
\left(\frac{R_{\star}}{r}\right)^{n_1}\frac{\left(r-R_{\star}\right)^{n_2}}{\left(r_{b}-R_{\star}\right)^{n_2}}; & $r>r_b$.\cr
}
\label{rho4}
\end{eqnarray}
\noindent with $r_{b}=10^{12}$ cm,  $R_{\star}\simeq3\times10^{12}$ cm, $n_1=17/7$ and $n_2=5$.\\
\section{Thermal Neutrino expectation}
\label{section3}
During the hypercritical accretion episode a copious amount of neutrinos in the energy range  of 1 - 30 MeV is produced. The processes of neutrino emission involved in the episode are:
\begin{itemize}
\item Electron-positron annihilation process.  $e^{\pm}$ pairs are annihilated to create neutrino-antineutrino pairs  ($e^{-}+e^{+}\rightarrow\nu+\bar{\nu}$).
\item Photo-neutrino process.  A  photon is Compton scattered to produce a neutrino-antineutrino pair ($\gamma+e^{\pm}\rightarrow e^{\pm}+\nu+\bar{\nu}$).
\item Plasmon decay process. A photon propagating within an electron plasma (plasmon) is spontaneously transformed into a neutrino-antineutrino pair ($\gamma\rightarrow\nu+\bar{\nu}$).
\item Bremsstrahlung process. A neutrino-antineutrino pair is created by either an electron-nucleon interaction ($e^{\pm}+N\rightarrow e^{\pm}+N+\nu+\bar{\nu}$) or nucleon-nucleon interaction ($N+N\rightarrow N+N+\nu+\bar{\nu}$).
\end{itemize}
\subsection{Number of  Expected Neutrinos}
The expected number of neutrinos can be estimated requiring the neutrino luminosity of the  hypercritical phase $L=4\pi d^2_z   E^2 dN/dE$ and the effective volume of the neutrino detector $V$.  The expected event rate can be written as 
\be\label{rate}
N_{ev}=V N_A\,  \rho_N \int_t \int_{E'} \sigma^{\bar{\nu}_ep}_{cc} \frac{dN}{dE}\,dE\, dT\,,
\ee

\noindent where  $N_A=6.022\times 10^{23}$ g$^{-1}$ is the Avogadro's number, $\rho_N=2/18\, {\rm g\, cm^{-3}}$ is the nucleons density in water \citep{2004mnpa.book.....M}, $ \sigma^{\bar{\nu}_ep}_{cc}\simeq 9\times 10^{-44}\,E^2_{\bar{\nu}_e}/MeV^2$  is the cross section \citep{1989neas.book.....B} and $dT$ is the detection time of neutrinos. The number of events (eq. \ref{rate}) is
\bary\label{num_Neu}
N_{ev}&\simeq&\frac{t}{<E_{\bar{\nu}_e}>}V N_A\,  \rho_N  \sigma^{\bar{\nu}_ep}_{cc} <E_{\bar{\nu}_e}>^2<\frac{dN}{ dE_{\bar{\nu}_e}}>\cr
&\simeq&\frac{t}{4\pi d^2_z <E_{\bar{\nu}_e}>}V N_A\,  \rho_N  \sigma^{\bar{\nu}_ep}_{cc}\,L_{\bar{\nu}_e}\,,
\eary
where we have assumed an average energy  $<E_{\bar{\nu}_e}>\simeq E_{\bar{\nu}_e}$ during a period of  time $t$ \citep{2004mnpa.book.....M}. 
\subsection{The Hyper-Kamiokande Experiment}
The Hyper-Kamiokande detector will be the third generation water Cherenkov in Kamioka, designed for a vast variety of neutrino studies.  At 8 km south of its predecessor Super-Kamiokande,  the Hyper-Kamiokande detector will be located  in the Tochibora mine of the Kamioka Mining and Smelting Company, near Kamioka town in the Gifu Prefecture, Japan.   This detector consists of two separate caverns, each having an egg-shaped cross section of 48 meters wide, 54 meters tall, and 250 meters long. The entire array consists of 99.000 PMTs, uniformly surrounding the region. It will have a total (fiducial) mass of 0.99 (0.56) million metric tons, approximately 20 (25) times larger than that of Super-Kamiokande.  Among the physical potentials of this detector is the detection of astrophysical neutrinos from the Supernova remnant and the studies of neutrino oscillation parameters \citep{2011arXiv1109.3262A, 2014arXiv1412.4673H}.
\subsection{Neutrino Oscillation}
The  properties of these  neutrinos are modified when they propagate in matter.   In the following subsections we will describe the dynamics of neutrinos going through matter.
\subsubsection{Two-Neutrino Mixing}
The neutrino oscillations for two-neutrino mixing ($\nu_e\leftrightarrow \nu_{\mu, \tau}$) are determined by the evolution equation in matter given by 
{\small
\be
\frac{d}{dt}{\pmatrix{
\nu_{e} \cr
\nu_{\mu}
}}
=-i \frac{\delta m^2}{4E_{\nu}}
{\pmatrix{
\frac{4E_\nu V_{eff}}{\delta m^2}-2 \cos 2\theta & \sin 2\theta \cr
\sin 2\theta  & 0
}}
{\pmatrix{
\nu_{e} \cr
\nu_{\mu}
}}\,,
\ee
}
where $V_{eff}$ is the neutrino effective potential generated by the medium,   $E_{\nu}$ is the neutrino energy,  $\delta m^2$ is the neutrino mass difference and $\theta$ is the neutrino mixing angle.    From  the conversion probability $P_{\nu_e\rightarrow {\nu_{\mu}{(\nu_\tau)}}}(t) = (\frac{\delta m^2}{2\omega E_\nu} \sin (2\theta)\,\sin\left (\omega t/2\right))^2$ with $\omega=\frac{\delta m^2}{2E_\nu} \,\sqrt{(\frac{2E_\nu V_{eff}}{\delta m^2} - \cos 2\theta)^2+ \sin^2 2\theta}$, the oscillation length of neutrino is written as
\be
L_{osc}=\frac{L_v}{\sqrt{(\frac{2E_\nu V_{eff}}{\delta m^2} - \cos 2\theta)^2+ \sin^2 2\theta}},
\label{osclength}
\ee
where $L_v=4\pi E_\nu/\delta m^2$ is the oscillation length in vacuum. The resonance condition and resonance length can be written as
\be
V_{eff}=\frac{\delta m^2}{2E_\nu} \cos 2\theta\,,
\label{reso2d}
\ee
and%
\be
L_{res}=\frac{L_v}{\sin 2\theta},
\label{oscres}
\ee
respectively. 
\subsubsection{Three-neutrino Mixing}
The dynamics of neutrino oscillations in matter for three-neutrino mixing is determined by 
\be
\frac{d\vec{\nu}}{dt}= -i [U\cdot H^d_0\cdot U^\dagger+diag(V_{eff},0,0)] \vec{\nu},
\ee
where $H^d_0=\frac{1}{2E_\nu}diag(-\delta m^2_{21},0,\delta m_{32}^2)$, $U$ is the matrix of three-neutrino mixing \citep{gon03,akh04,gon08,gon11} and the state vector in the flavor basis is $\vec{\nu}\equiv(\nu_e,\nu_\mu,\nu_\tau)^T$. The survival and conversion  probabilities  for electron, muon and tau are given in \cite{2014ApJ...787..140F}. The oscillation length of neutrino for three-neutrino mixing is given by
\be
L_{osc}=\frac{L_v}{\sqrt{\cos^2 2\theta_{13} (1-\frac{2 E_{\nu} V_{eff}}{\delta m^2_{32} \cos 2\theta_{13}} )^2+\sin^2 2\theta_{13}}},
\label{osclength}
\ee
where $L_v=4\pi E_{\nu}/\delta m^2_{32}$.  The resonance condition and resonance length are given by
\be
1- \frac{2 E_{\nu} V_{eff}}{\delta m^2_{32} \cos 2\theta_{13}} =0\,,
\label{reso3d}
\ee
and%
\be
L_{res}=\frac{L_v}{\sin 2\theta_{13}},
\label{oscres}
\ee
respectively. 
\subsubsection{Neutrino Oscillation from  Source to Earth}
Neutrinos oscillate in their paths to Earth.  In this case, the dynamics is governed  by the Schr$\ddot{o}$dinger equation in vacuum and the Hamiltonian is given as a function of neutrino mixing parameters. The conversion probability from a flavor state $\alpha$ to a flavor state $\beta$ is written  as {\small $P_{\nu_\alpha\to\nu_\beta} =\mid <  \nu_\beta(t) | \nu_\alpha(t=0) >  \mid =\delta_{\alpha\beta}-4 \sum_{j>i}\,U_{\alpha i}U_{\beta i}U_{\alpha j}U_{\beta i}\,\sin^2\left(\delta m^2_{ij} L/4\, E_\nu \right)$}, where  $U_{j i}$ is once again the three neutrino  mixing \citep{gon03,akh04,gon08,gon11}. Here  the sin term in the probability has been averaged  to $\sim 0.5$ for distances (L) longer than the Solar System \citep{lea95}. 
\section{Results}
\label{section4}
\subsection{Plasma-dynamics on the stellar surface}
We carry out hydrodynamic numerical simulations of the hypercritical accretion phase in a two-dimensional spherical mesh, using the Flash method developed by  \citet{Fryxell2000}. This multi-physics Eulerian and parallelized code was designed to figure out many problems in high-energy astrophysics, especially those  related to the dynamics of plasmas around neutron stars. Recently, using a customized version of such numerical code, \cite{2014MNRAS.442..239F} presented numerical simulations of strong  matter accretion  onto the newborn neutron star to compute the size of the neutrino-sphere, the emissivity and the luminosity of neutrinos in the SN1987A scenario.  Requiring adequate parameters to describe the Cas A scenario, we will follow  a similar procedure to that performed for SN1987A.\\
The Flash code uses the piecewise-parabolic method (PPM) solver  to solve the whole set of hydrodynamic equations for a compressible gas dynamics in one, two, or three spatial dimensions. This directionally-split method makes use of a second-order Strang time splitting, which uses a finite-volume spatial discretization of the Euler equations together with an explicit forward time difference. Time-advanced fluxes at cell boundaries are computed using the numerical solution to a Riemann shock-tube problem in each boundary. Initial conditions for each Riemann problem are determined by assuming the non-advanced solution to be piecewise-constant in each cell. Requiring the Riemann solution, the effect of explicit nonlinearity is introduced into the differential equations. It allows the calculation of sharp shock fronts and contact discontinuities without introducing significant nonphysical oscillations into the flow. Since the value of each variable in each cell is assumed to be constant, this method is limited to first-order accuracy in both space and time. PPM improves on Godunovs method by representing the flow variables with piecewise-parabolic functions. It also uses a monotonicity constraint rather than artificial viscosity to control the oscillations near discontinuities. The Euler equations can be written in conservative form as
\begin{eqnarray}
\frac{\partial \rho }{\partial t}+\mathbf{\nabla }\cdot \left( \rho \mathbf{v%
}\right) &=&0 \label{mass} \\
\frac{\partial \rho \mathbf{v}}{\partial t}+\mathbf{\nabla }\cdot \left(
\rho \mathbf{vv}\right) +\nabla P &=&\rho \mathbf{g}\label{momentum} \\
\frac{\partial \rho E}{\partial t}+\mathbf{\nabla }\cdot \left[\mathbf{v}%
\left(\rho E+P\right)\right]%
&=&\rho \mathbf{v\cdot g}\label{energy}\,,
\end{eqnarray}

\noindent where the previous equations (\ref{mass}, \ref{momentum} and \ref{energy}) represent the conservation of mass, momentum and total energy, respectively. In this context, $\rho$ is the fluid density, $\mathbf{v}$ is the fluid velocity, $P$ is the pressure, $\mathbf{g}$ is the acceleration due to gravity, $t$ is the time coordinate and $E$ represents the sum of the internal energy $\epsilon$ and kinetic energy per unit mass
\be
E=\epsilon+\frac{1}{2}|\mathbf{v}|^{2}.
\ee

\noindent The pressure is obtained from the energy and density using the equation of state. In this case, we use a customized version of the Helmholtz Unit of the Flash code which includes contributions from the nuclei, $e^\pm$ pairs, radiation and the Coulomb correction. This routine is appropriate for addressing astrophysical phenomena in which electrons and positrons may be relativistic and/or degenerated, and the radiation may significantly contribute to the thermodynamic state. The pressure and internal energy are calculated considering all components
\begin{eqnarray}
P_{tot} &=&P_{rad}+P_{ion}+P_{ele}+P_{pos}+P_{coul}\,, \\
\epsilon _{tot} &=&\epsilon _{rad}+\epsilon _{ion}+\epsilon
_{ele}+\epsilon _{pos}+\epsilon _{coul}\,.
\end{eqnarray}

\begin{figure*}
\centering
\includegraphics[width=\textwidth]{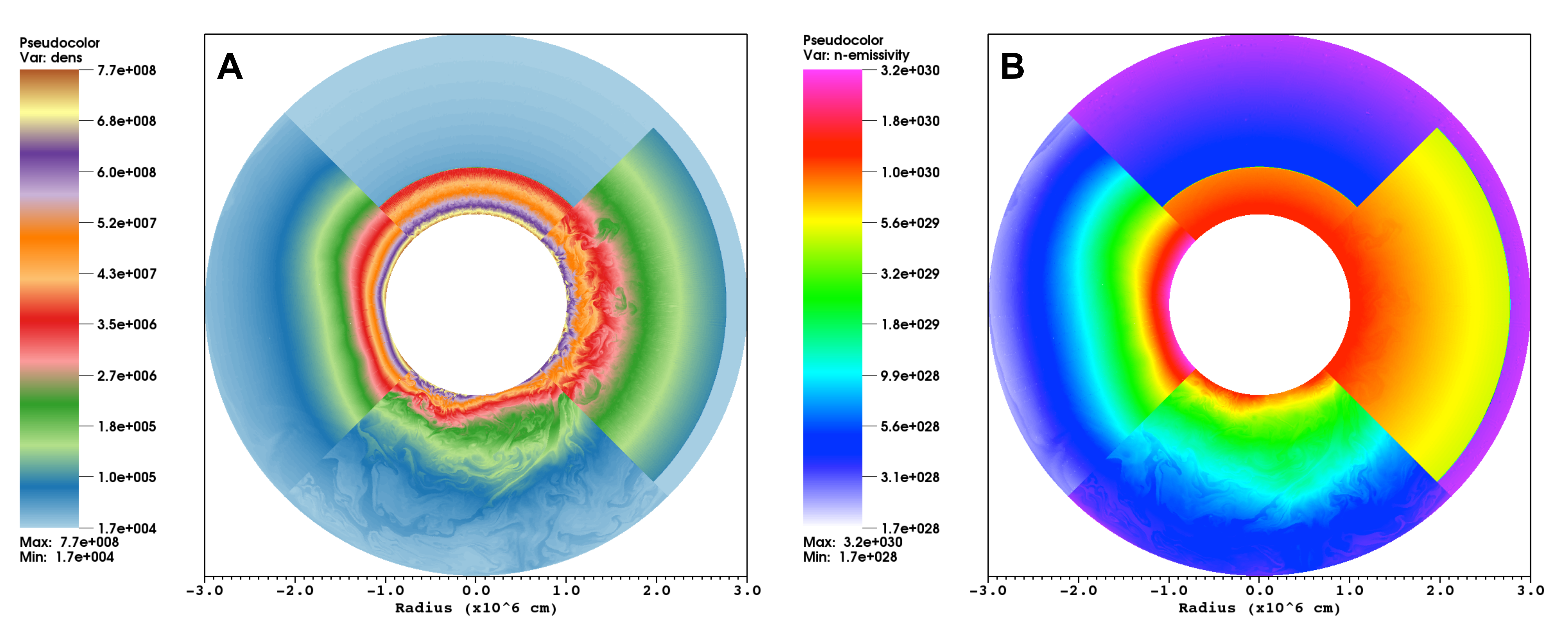}
\caption{\label{Fig-Gio1} Pseudo-color of density (A) and neutrino emissivity (B), for the Cas A parameters: the initial transient, $t=0.2$ ms (up); shock evolving in the domain, $t=1$ ms (right); shock leaves the domain and the transient begins to disappear, $t=5$ ms (down); and a quasi-hydrostatic equilibrium envelope is formed with a new crust on the stellar surface, $t=30$ ms (left).}
\end{figure*}

\begin{figure*}
\centering
\includegraphics[width=0.8\textwidth]{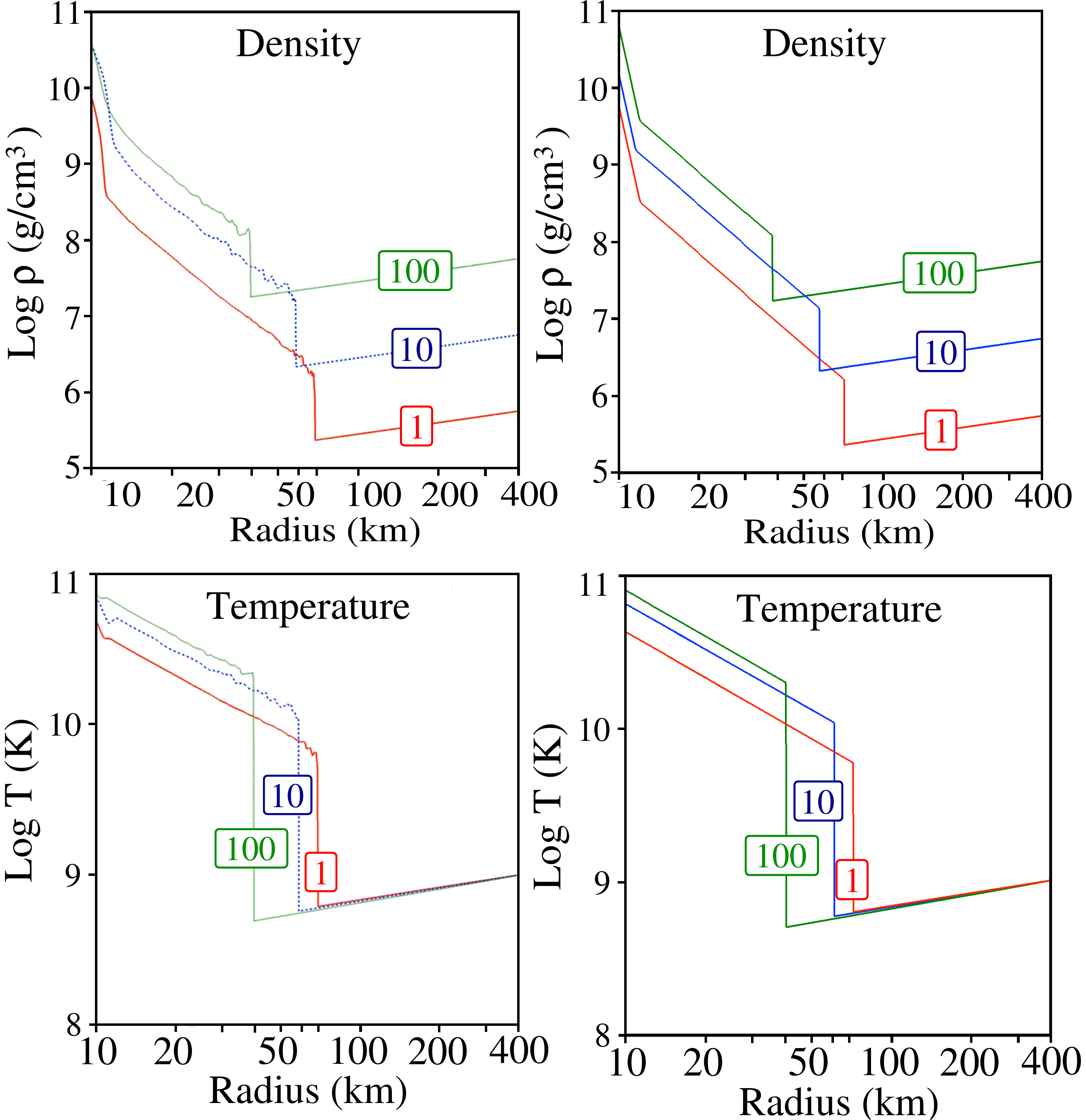}
\caption{Density and temperature profiles as a function of the stellar radius. In the left-hand panels we show the density and temperature profiles obtained with MHD simulations with similar parameters of Cas A (\citet{Bernal2013}), whereas in the right-hand panels the analytical approach developed in \citet{Chevalier1989} is shown. Here, [1] represents the Cas A accretion rate, and [10] and [100] are accretion rates one and two orders of magnitude larger, respectively. Note the new crust formed in the hypercritical regime very close to the neutron star surface ($\sim 300$ m). The quasi-hydrostatic envelope lies between the new crust until $\sim 50$ km. The free-fall region is the remaining region, for the Cas A parameters. \label{Fig-densities}}
\end{figure*}

\begin{figure*}
\centering
\includegraphics[width=\textwidth]{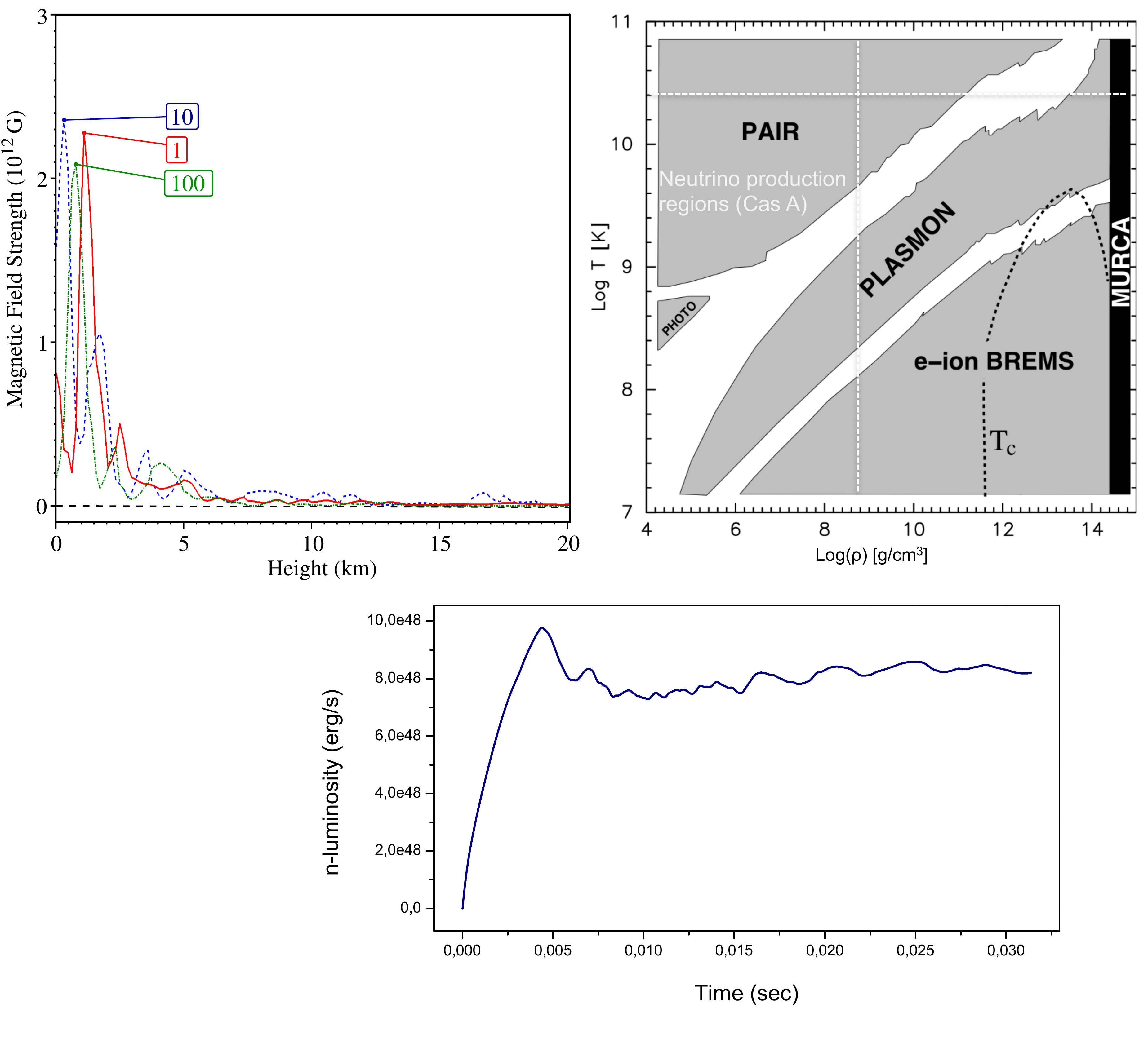}
\caption{Left-hand panel above:  Magnetic Field Profiles as a function of radius are plotted. Note the complete submergence of the magnetic field at the same height scale of the neutrino-sphere. Here, [1] represents the Cas A accretion rate, and [10] and [100] are accretion rates one and two orders of magnitude larger, respectively. Right-hand panel above:  Parameter space of temperature and density, and regions where  each neutrino cooling process is dominant. The white-dashed lines divide the region of neutrino production for Cas A.  Panel below:   We show the neutrino luminosity as a function of time for the hydrodynamic case. This plot was obtained integrating over the entire computational domain and using the tabulated numerical values performed by \citet{Itoh1996} \label{Fig-regions}.}
\end{figure*}

\noindent Here the subscripts \textquotedblleft rad\textquotedblright , \textquotedblleft ion\textquotedblright , \textquotedblleft ele\textquotedblright , \textquotedblleft pos\textquotedblright\ and \textquotedblleft coul\textquotedblright\ represent the contribution from radiation, nuclei, electrons, positrons, and corrections for Coulomb effects, respectively. The radiation portion assumes a blackbody in local thermodynamic equilibrium, the ion portion (nuclei) is treated as an ideal gas with $\gamma =5/3,$ and the electrons and positrons are treated as a noninteracting Fermi gas. That is, the number densities of
free electrons $N_{ele}$ and positrons $N_{pos}$ in the noninteracting Fermi gas formalism are given by
\be
N_{ele}=\frac{\sqrt{2}}{\pi^2}m_{e}^{3}\beta^{3/2}\left[F_{1/2}(\eta_{ele},\beta)+F_{3/2}(\eta_{ele},\beta)\right]\,,
\ee
\be
N_{pos}=\frac{\sqrt{2}}{\pi^2}m_{e}^{3}\beta^{3/2}\left[F_{1/2}(\eta_{pos},\beta)+\beta F_{3/2}(\eta_{pos},\beta)\right]\,,
\ee
\noindent where $m_e$ is the electron rest mass, $\beta=T/m_{e}$ is the normalized temperature, $\eta_{ele}=\mu/T$ is the normalized chemical potential for electrons, $\eta_{pos}=-\eta_{ele}-2/\beta$ is the normalized chemical potential for positrons, and $F_{k}(\eta,\beta)$ is the Fermi-Dirac integral given by
\be
F_{k}(\eta,\beta)=\intop_{0}^{\infty}\frac{x^{k}(1+0.5\beta x)^{1/2}dx}{\exp(x-\eta)+1}\,.
\ee
\noindent Because the electron rest mass is not included in the chemical potential, the positron chemical potential must have the form given by $\eta_{pos}$. For complete ionization, the number density of free electrons in the matter is
\be
N_{ele,matter}=\frac{\overline{Z}}{\overline{A}}N_{a}\rho=\overline{Z}N_{ion},
\ee
\noindent and charge neutrality requires
\be
N_{ele,matter}=N_{ele}-N_{pos}.
\ee
\noindent Solving the previous equations with a standard one-dimensional root-finding algorithm we determine $\eta$, and the Fermi-Dirac integrals can be evaluated given the pressure, specific thermal energy, and entropy due to the free electrons and positrons. Full details of the Helmholtz equation of state are provided in \citet{2000ApJS..126..501T}.\\
As was pointed out, neutrino energy losses are dominated by the annihilation processes which involve the formation of neutrino-antineutrino pairs through the  $e^{\pm}$ annihilation near the stellar surface ($e^{+}+e^{-}\rightarrow\nu+\bar{\nu}$). In addition, we include other relevant neutrino processes present in such regime, which are implemented in a customized module and described in \cite{Itoh1996}. We ignore neutrino absorption and heating, and nuclear reactions. \\
Recently, \citet{Bernal2010, Bernal2013} showed that, regardless of its initial configuration, the magnetic field is buried and submerged under the stellar surface during the hypercritical phase.  Therefore, as the magnetic field is submerged within the surface, it does not play an important role in the dynamics of quasi-hydrostatic envelope, although the dynamics itself of the envelope is relevant.  Here, we do not take into account the magnetic field and just focus on the hydrodynamics near the stellar surface. The hydrodynamic numerical simulations of the hypercritical accretion regime are carried out in a two-dimensional spherical mesh $\left(r,\theta\right)$, where the radial component $r$ lies in the range $10^{6}\, {\rm cm} \leq r \leq 3\times10^{6}\,{\rm cm}$. Due to the symmetry of the problem, we only simulate a quarter of the total domain $(\pi/4\leq\theta\leq3\pi/4)$ which has $2048\times2048$ effective zones as spatial resolution. It is worth noting that it corresponds to the maximum level resolution which is equal to 6 for this work. The time resolution of these simulations is $dt\simeq10^{-7}$~s. We use, for the discretization grids, a block-structured oct-tree based on an adaptive grid used by the PARAMESH library which is included in FLASH code. If the Adaptive Mesh Refinement (AMR) grid is used, then the formation of the physical domain starts at the lowest level of refinement. Initial conditions are applied to each block at this level by calling initial condition routines. The Grid Unit then checks the refinement criteria in the blocks that it has created and also refines these blocks when the criteria are met. It then calls  the routines of the initial conditions to initialize the newly created blocks. This process repeats until the grid reaches the required refinement level in the areas marked for refinement.\\
In the boundary conditions, we impose mass inflow along the top edge of the computational domain. We assume that the accreting matter is not magnetized. Concerning the FLASH code, the parallel structure of blocks means that each processor works independently. If a block is on a physical boundary, the guard cells are filled by calculations since there are not any neighboring blocks from which to copy values. At the top of the domain, initially we set the velocity   in free fall to be $v_{r}=\sqrt{2GM/r}$ in all the guard cells, and set a density profile which fixes a constant accretion rate, $\rho =\dot{M}/{4\pi r^{2}v_{r}}$, as in the computational domain. The temperature is uniform and set to $T=10^{9}$ K.  Other variables are calculated with the equation of state.  When the shock reaches this boundary, we allow it to leave the computational domain. At the same time, this boundary switches to the Quasi-hydrostatic Envelope mode, i.e, it is adaptable for consistency with the model. For this purpose we use the $\rho$, $p$, and $v$ profiles given by the analytical solution (eqs. \ref{equation2} and \ref{rs}).  It is justified due to the  strong initial transient and short duration of the hypercritical phase. 
At the bottom, on the neutron star surface, we use a custom boundary condition that enforces the hydrostatic equilibrium. In order to establish this boundary in the problem of hypercritical accretion, we fix the velocities as null in all the guard cells, $(v_{r}=v_{\theta}=0),$ and copy the density and the pressure of the first zone of the numerical domain. This zone corresponds to the neutron star surface $\left( \rho =\rho(ic), p=p(ic)+\rho v^{2}+\rho gh\right)$, where $ic$ is the first zone in the domain. The rest of thermodynamics variables are calculated from the equation of state. In addition, we have implemented periodic conditions along the sides. Periodic (wrap-around) boundary conditions are initially configured in this routine as well. If periodic boundary conditions are set in the x-direction, for instance, the first blocks in the x-direction are set to have as their left-most neighbor the blocks that are the last in the x-direction, and vice versa. Thus, when the guard cell filling is performed, the periodic boundary conditions are automatically maintained.\\
We focus our study on the dynamics of the initial transient when the reverse shock rebounds off the stellar surface and builds a quasi-hydrostatic equilibrium envelope, instead of following the shock. We start the simulation requiring the free-fall profiles described in subsection \ref{FF-zone} and considering an initial temperature of $10^{9}$ K. The  initial value of temperature was elected because it is close to the value expected from the theoretical model, and then the code finds the correct pressure profile after some steps of simulation. The neutron star mass and the accretion rate, for the Cas A parameters, are chosen as $M\simeq1.4 ~{\rm M_\odot}$ and $\dot{M}\simeq 500 ~{\rm M_{\odot}}\, {\rm yr^{-1}}$, respectively. Fig. \ref{Fig-Gio1} shows the time evolution color maps of density (A) and neutrino emissivity (B) for different times: $t=0.2$ ms (up) shows the initial transient of the third expansive shock with material falling onto it; $t=1$ ms (right) illustrates the evolution of the shock in the computational domain resulting in a complex morphology above the stellar surface; $t=5$ ms (down) exhibits the moment when the shock reaches the external boundary and leaves the computational domain; and finally, $t=30$ ms (left) shows the quasi-hydrostatic envelope established with a new thin crust on the neutron star surface. The hydrodynamic instabilities and the rich morphology observed inside the envelope in early times have disappeared. Note that in the new thin crust the neutrino cooling is very effective creating an energy sink that allows the material to be deposited on the surface slowly. That is, in such region the neutrino emissivity is highly efficient.\\
In this analysis, not only  hydrodynamic simulations of the hypercritical regime are performed, but also the results of previous MHD simulations for similar parameters of Cas A are presented.  Such results show a complete magnetic field submergence under the new formed crust, as it was reported in \citet{Bernal2010, Bernal2013}. In these papers, to perform the MHD simulations the authors used the Split Eight-Wave solver to solve the whole set of MHD equations. They considered wide accretion columns in 2D and 3D to follow the accretion shock and also a magnetic field loop (in the shape of an hemi-torus) as magnetic initial condition. On the central hemi-circle of the loop the field had a strength $B_0 = 10^{12}$ G,  it was shaped as a Gaussian, i.e., with strength $B(d) = B_0 \exp\left[-(d/R_L)^2\right)]$, $d$ being the distance to the loop central hemi-circle and $R_L = 1$ km. The two feet of the loop were centered at $x=-5$ and $x=+5$ km, and $z=0$ in the 3D model. When magnetic field was present, it was put at the bottom boundary in such a form that it was continuous from the guard cell to the physical domain, i.e, the authors anchored the magnetic field onto the neutron star surface and in the rest of the guard cells it was null. On the other hand, Figure \ref{Fig-densities} shows radial profiles of density and temperature as a function of the stellar radius when the envelope has been well established. It is worth noting that density profiles obtained by MHD and HD are very similar, hence we show only those densities and temperatures obtained by MHD. This is justified because in this case we were able to follow the evolution of the shock until it formed the quasi-hydrostatic envelope with matter falling over it. In the left-hand panels we show the density and temperature profiles obtained from simulations for several hyperaccretion rates, in terms of a fiducial accretion rate ($\dot{M}\simeq 500 ~{\rm M_{\odot}}\, {\rm yr^{-1}}$), whereas in the right-hand panels we show the analytical solutions (with the same rates), using the analytical model  \citep{Chevalier1989}. Note the excellent agreement between them. The three regions (new crust, envelope and free fall) are visible and distinct. In this case, label [1] represents the Cas A accretion rate, and labels [10] and [100] are accretion rates one and two orders of magnitude larger, respectively. The envelope is well established in 600 ms for label [1], 300 ms for label [10] and 100 ms for label [100].
A detail comparison between the MHD and HD solvers was reported in \citet{Bernal2010}.  In the new crust zone,  the magnetic field is compressed and amplified, as shown in Figure \ref{Fig-regions} (upper-left panel). The magnetic field submergence into the new crust formed in the hypercritical regime allows us to justify the fact that the magnetic field does not play an important role in the subsequent dynamics of the envelope. 
Both in the hydrodynamic and MHD cases the estimated size of the neutrino-sphere (including all the relevant neutrino cooling processes) is $r\simeq3.7\times10^{5}\:\mathrm{cm}\simeq(1/3)R$. The mean value of emissivity in this region is $\dot{\epsilon}_{\nu}\simeq2.4\times10^{30}\:\mathrm{erg\, s^{-1}\, cm^{-3}}$. The volume of the neutrino-sphere is given by $V\simeq(4/3)\pi R^{3}=(4/3)\pi\times10^{18}\:\mathrm{cm^{3}}$ and then the neutrino luminosity is obtained to be $L_{\nu}\simeq8.4\times10^{48}\:\mathrm{erg\, s^{-1}}$. This estimate agrees with the numerical value obtained by integrating directly the neutrino luminosity in the whole computational domain. The upper-right panel in Figure. \ref{Fig-regions} shows the parameter-space of temperature and density. Taking into account the \citet{Itoh1996} tables, this panel has been divided into grey zones, highlighting the different regions where neutrino cooling processes dominate. Also, we have drawn lines in white color to show the region of interest for the Cas A parameters. The panel below in Figure \ref{Fig-regions} shows the time evolution of the neutrino luminosity integrated in the whole computational domain for the hydrodynamic case. In the initial transient the luminosity reaches a maximum value, showing then small oscillations about a fixed value. Similar results are found in the MHD case.

\subsection{Neutrino expectation}
\begin{figure*}
\centering
\includegraphics[width=\textwidth]{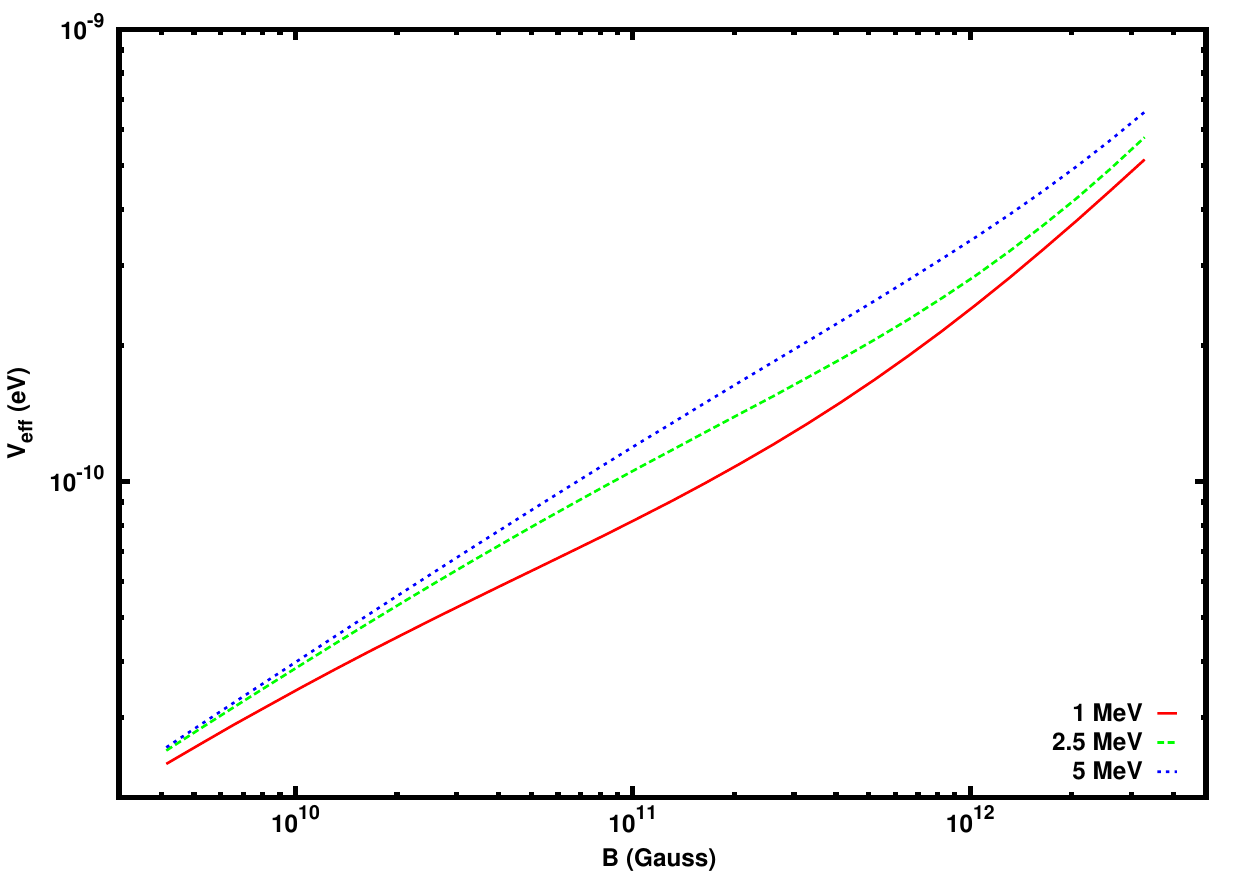}
\caption{Neutrino effective potential is plotted as a function of magnetic field for three values of temperatures (1, 2.5 and 5 MeV). We have used the neutrino energy of 5 MeV \label{potential}}
\end{figure*}

\begin{figure*}
\centering
\includegraphics[width=\textwidth]{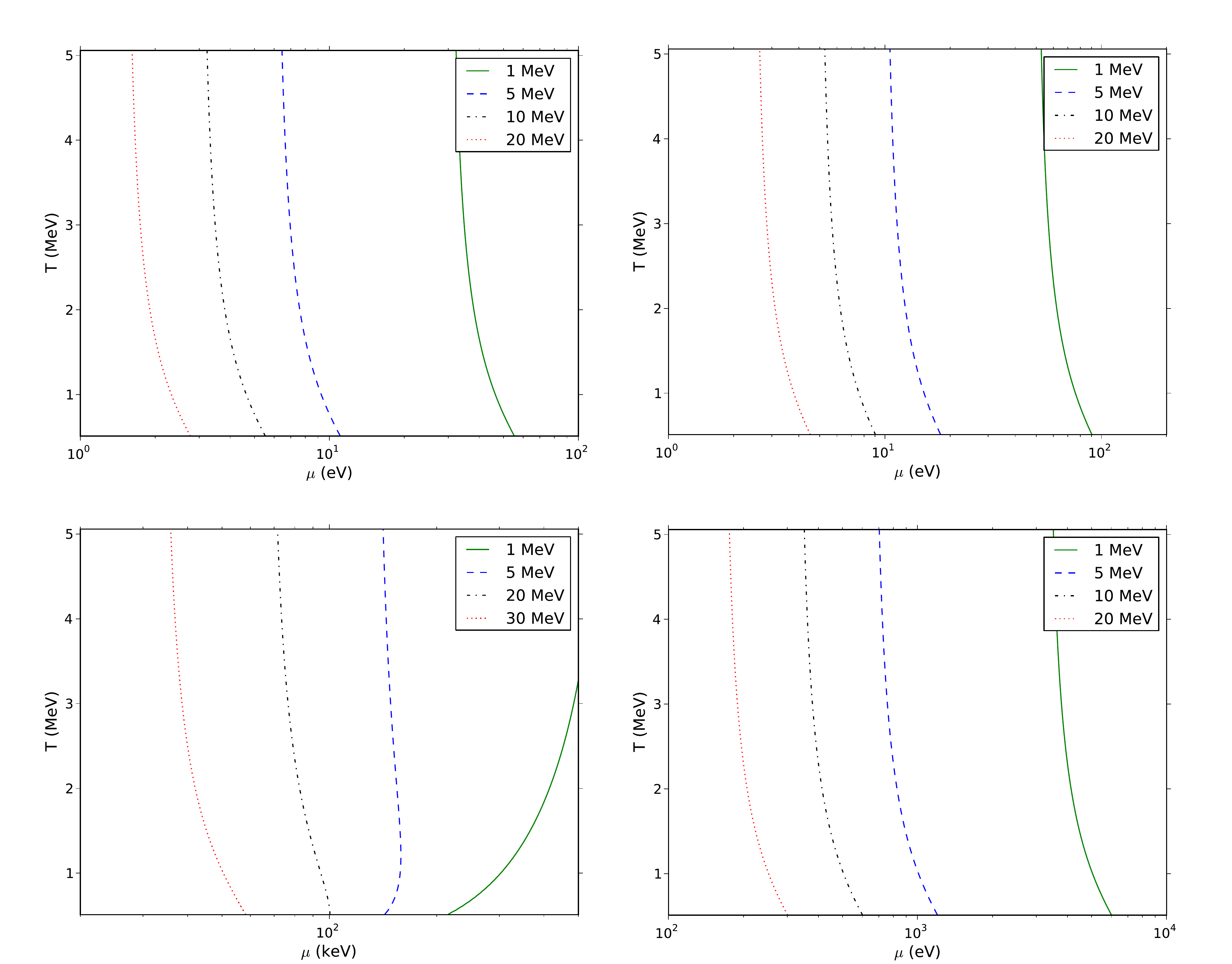}
\caption{Contour plots of temperature and chemical potential as a function of neutrino energy for which the resonance condition is satisfied.   We have required the best-fit values of the two-neutrino mixing  (solar, top left; atmospheric, top right; and accelerator, bottom left) and three-neutrino mixing (bottom right).\label{res_W}  \label{oscillation}}
\end{figure*}

\begin{figure*}
\centering
\includegraphics[width=0.96\textwidth]{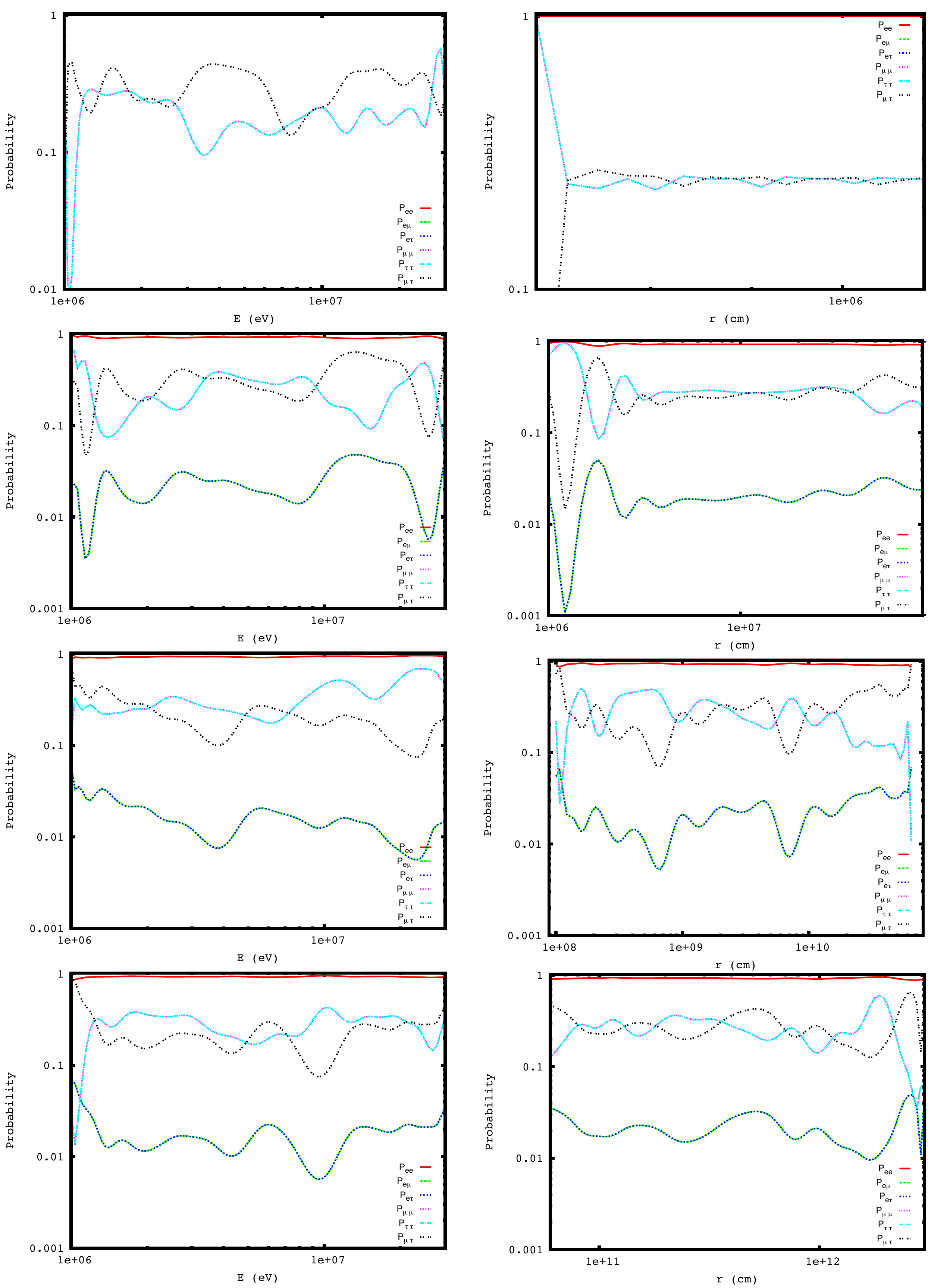}
\caption{Oscillation probabilities of neutrinos as a function of neutrino energy (left) and distance (right) for the three-neutrino mixing parameters. We show  the four regions (from top to bottom panels): The new crust on the neutron star surface (top), quasi-hydrostatic envelope (second), free fall zone (third) and external layers (four). \label{probabilities}}
\end{figure*}

Taking into account the values of neutrino Luminosity  $L_{\bar{\nu}_e}=8.4\times10^{48}\:\mathrm{erg\, s^{-1}}$, effective volume $V\simeq 0.56\times 10^{12}\,{\rm cm^3}$ \citep{2014arXiv1412.4673H}  and the average neutrino energy  $<E_{\bar{\nu}_e}>\simeq 15\, {\rm  MeV}$, from eq. (\ref{num_Neu}) we get that the number of events expected from the hypercritical accretion episode on Hyper-Kamiokande is 3195.\\
These thermal neutrinos will oscillate when they propagate through the star.  On the neutron star surface,  the plasma is highly magnetized with B$\simeq3\times 10^{12}$ G (see fig.  \ref{Fig-regions} left-hand panel) and  thermalized at T $\simeq$ 3 MeV (see fig.  \ref{Fig-densities}  panels below).  The neutrino effective potential at the moderate field limit is given by \citep{2014ApJ...787..140F}
{\small
\begin{eqnarray}
V_{eff}=\frac{\sqrt2 G_F \,m_e^3\,B}{\pi^2 B_c} \biggl[\sum^\infty_{l=0} (-1)^l \sinh\alpha_l\,\biggl\{\frac{m_e^2}{m^2_W}\biggl(1+4\frac{E^2_\nu}{m^2_e}\biggr)K_1(\sigma_l)\cr
+\sum^{\infty}_{n=1}\lambda_n\,\biggl( 2+\frac{m_e^2}{m^2_W}\biggl( 3-2\frac{B}{B_c}+4\frac{E^2_\nu}{m_e^2}\biggr)\biggr)\, K_1(\sigma_l\lambda_n) \biggr\}\cr \nonumber
-4\frac{m_e^2}{m^2_W}\frac{E_\nu}{m_e}\sum^\infty_{l=0} (-1)^l \cosh\alpha_l\biggl\{ \frac34K_0(\sigma_l)+\sum^{\infty}_{n=1}\lambda^2_n\, K_0(\sigma_l\lambda_n)  \biggr\}\biggr],
\label{fpoteff}
\end{eqnarray}
}
where K$_i$ is the modified Bessel function of integral order i, $\mu$ is the chemical potential,  $\lambda^2_n=1+2\,n\,B/B_c$,    $\alpha_l=(l+1)\,\mu/T$, $\sigma_l=(l+1)\,m_e/T$ and $B_c=4.4\times 10^{13}$ G.  As shown in Figure \ref{potential}, the effective potential is an increasing function of magnetic field for T=1 MeV, 2.5 MeV and 5 MeV. The value of the effective potential lies in the ranges: $2.3\times 10^{-11} {\rm eV}$  - $ 5.1\times 10^{-10} {\rm eV}$ for T=1 MeV,  $2.61\times 10^{-11} {\rm eV}$  - $5.7\times 10^{-10} {\rm eV}$ for 2.5 TeV, and  $2.64\times 10^{-11} {\rm eV}$ - $6.3\times 10^{-10} {\rm eV}$ for 5 MeV when the magnetic field lies at  $4\times 10^{9}\, {\rm G} \leq B \leq 4\times 10^{12}\, {\rm G}$.  From this figure, one can see  that the effective potential is positive, therefore  due to the positivity of the effective potential ($V_{eff}>$ 0) neutrinos can oscillate resonantly. Taking into account the parameters of the two (table 2) and three-neutrino (table 3) mixing we analyze the resonance condition, as shown in Figure \ref{oscillation}. 
\begin{table}
\begin{center}\renewcommand{\tabcolsep}{0.2cm}
\renewcommand{\arraystretch}{1.1}
\begin{tabular}{ccccc}\hline
{\small Parameters}  & $\delta m^2 ({\rm eV^2}$)  & angle \\ \hline
{\small Solar}     &  ${\small  5.6^{+1.9}_{-1.4}\times 10^{-5}\, }$  &  ${\small \tan^2\theta=0.427^{+0.033}_{-0.029}}$   \\\hline
{\small Atmospheric}   & ${\small  2.1^{+0.9}_{-0.4}\times 10^{-3} }$  &  ${\small \sin^22\theta=1.0^{+0.00}_{-0.07}}$ \\\hline
{\small Accelerator}  & ${\small 0.5\,{\rm eV^2}}$  &  ${\small \sin^2\theta=0.0049}$  \\\hline
\end{tabular}
\end{center}
\caption{The best fit values of two-neutrino mixing (solar, atmospheric and accelerator neutrino experiments) \citep{aha11, abe11a, ara05, shi07,mit11}. }
\label{flaratio_s}
\end{table}
\begin{table}
\begin{center}\renewcommand{\tabcolsep}{0.2cm}
\renewcommand{\arraystretch}{0.6}
\begin{tabular}{cccl}\hline
{\small Parameters}  & {\small Best fit} \\ \hline
${\small  \delta m^2_{21} }$    &  ${\small  7.62 \times 10^{-5}\, {\rm eV^2}}$    \\\hline
${\small  \delta m^2_{31}} $  & ${\small  2.55\times 10^{-3}\,  {\rm eV^2}}$   \\\hline
${\small \sin^2\theta_{12}}$   & ${\small  0.320 }$    \\\hline
${\small \sin^2\theta_{23}}$   & ${\small  0.613 }$    \\\hline
${\small \sin^2\theta_{13}}$   & ${\small  0.0246 }$    \\\hline
\end{tabular}
\end{center}
\caption{The best fit values of the three-neutrino mixing \citep{PhysRevD.86.073012}.}
\label{flaratio_s}
\end{table}
From this figure  we can see that temperature is a decreasing function of chemical potential,  for the values of temperature in the range 0.7 MeV $\leq T \leq$ 5 MeV. We found that the chemical potential is in the  range of  1.7 eV $\leq \mu \leq$ 32 eV for  solar (left-hand panel above),  2.8 eV $\leq \mu \leq$ 91 eV for atmospheric (right-hand panel above), 36 keV $\leq \mu \leq$ 465 keV for  accelerator (left-hand panel below) and  187 eV $\leq \mu \leq\,6.5\times 10^{3}$ eV, for three-neutrino (right-hand panel below) mixing. It can also be seen from fig. \ref{oscillation}  that the chemical potential achieves the largest value when accelerator parameters are considered, and the smallest one if solar parameters are taken into account.\\
Recently,  \citet{2014MNRAS.442..239F} showed that neutrinos can oscillate resonantly due to  the density profiles of the collapsing material surrounding the progenitor. The neutrino effective potential  associated  to each region is  $V_{eff}=\sqrt2 G_F N_A\,\rho(r) Y_e$  with $Y_e$ the number of electron per nucleon and $\rho(r)$ given by eqs. (\ref{rho2}), (\ref{rho3}) and (\ref{rho4}).   We plot the survival and conversion probabilities for the active-active ($\nu_{e,\mu,\tau} \leftrightarrow \nu_{e,\mu,\tau}$) neutrino oscillations in each region as shown in Figure \ref{probabilities}. From top to bottom panels, we show the oscillation probabilities when neutrinos pass through  the new thin crust on the neutron star surface (top), the quasi-hydrostatic envelope (second), the free fall zone (third) and the external layers (four). One can see that these probabilities vary with neutrino energy (left) and distance (right).  Taking into account the oscillation probabilities in each region and in the vacuum (on its path to Earth), we calculate the flavor ratio expected on Earth for four neutrino energies  ($E_{\nu}=5$ MeV, 10 MeV, 15 MeV and 20 MeV), as shown in table \ref{Table}. In this table we can see a small deviation from the standard ratio flavor 1:1:1.   In this calculation we take into account that for neutrino cooling processes (electron-positron annihilation, inverse beta decay, nucleonic bremsstrahlung and plasmons), only  inverse beta decay is the one producing electron neutrino. It is worth noting that our calculations of resonant oscillations were performed for neutrinos instead of anti-neutrinos, due to the positivity of the neutrino effective potential.\\
\begin{table*}      
\begin{center}
\caption[]{Here we show the neutrino flavor ratio in each zone of the collapsing star for  $E_{\nu}=$ 5, 10, 15 and 20 MeV.}\label{Table}
\begin{minipage}{126mm}
\begin{tabular}{lccccccccc}
\hline
 
 $E_{\nu}$  & On the NS surface & Accretion material & Free fall zone& Outer layers & On Earth \\\hline \hline 

{\small 5}     &  {\small 1.2:0.9:0.9}  &  {\small 1.176:0.912:	0.912} & {\small 1.151:0.925:0.925} &  {\small 1.124:0.938:0.938 }  &  {\small 1.039:0.978:	0.984} \\\hline

{\small 10}   & {\small 1.2:0.9:0.9}  &  {\small 1.160:0.920:0.920}&  {\small 1.130:0.935:0.935} &  {\small 1.113:0.944:	0.944 }  &  {\small 1.035:0.980:	0.986}\\\hline

{\small 15}  & {\small 1.2:0.9:0.9} &  {\small 1.158:0.921:0.921} &  {\small 1.132:0.934:0.934} &  {\small 1.129:0.936:0.936 }  &  {\small 1.040:0.977:	0.984}   \\\hline

{\small 20}  &  {\small 1.2:0.9:0.9}  &   {\small 1.177: 0.911:0.911}  &   {\small 1.147:0.926:0.926} &  {\small 1.146:0.927:0.927 }  &  {\small 1.045:0.973:0.982}  \\\hline
 
\end{tabular}
\end{minipage}
\end{center}
\end{table*}
Finally, we estimate the number of initial neutrino burst expected during the neutron star formation.  Considering a temperature T$\approx$ 4 MeV,  a duration of the neutrino pulse $t\simeq10$ s, the average neutrino energy  $<E_{\bar{\nu}_e}>\simeq 15\, {\rm  MeV}$  and a total fluence equivalent of Cas A  $\Phi\approx1.02 \times 10^{12}\, \bar{\nu}_e \,{\rm cm^{-2}}$ \citep{2004mnpa.book.....M, 1989neas.book.....B}, then the total number of neutrinos emitted from Cas A would be $N_{tot}=6\,\Phi\, 4\pi\, d_z^2\approx8.48\times 10^{57}$.  Similarly, we can compute the total radiated luminosity  corresponding to the binding energy of the neutron star $L_\nu\approx \frac{N_{tot}}{t}\times <E_{\bar{\nu}_e}>\approx 2\times 10^{52}$ erg/s. Taking into account the effective volume $V\simeq 0.56\times 10^{12}\,{\rm cm^3}$ \citep{2014arXiv1412.4673H}, from eq. (\ref{num_Neu}) we get  $N_{ev}\simeq2.29\times 10^6$ events that could have been expected during the neutron star formation in a neutrino detector as Hyper-Kamiokande experiment. Comparing  the number of neutrinos that would have been expected during the neutron star formation and  the hyper accretion phase, we would get  716.75 events.
\section{Conclusions}  
\label{section5}
Although the Cas A supernova event occurred 330 years ago, its similarity to other core-collapse supernovae allows to study the early evolution of the proto-neutron star inside the supernova remnant. In the present work, we have done a study of the hypercritical phase onto the newborn neutron star, in the Cas A scenario, in order to analyze the neutrino cooling effect when a quasi-hydrostatic envelope is formed. This phase is important because it allows to submerge the magnetic field within the new crust formed by the strong accretion. In this context, after the hyperaccretion has stopped, the magnetic field could diffuse back to the surface and result in a delayed switch-on of a pulsar \citep{1995ApJ...440L..77M}. Depending on the amount of accreted matter, the submergence could be deep enough that the neutron star may appear and remain unmagnetized for several centuries or millennia \citep{Geppert1999}.
Cas A has been classified as one of Central Compact Objects (CCO; \cite{Pavlov2004}) defined as X-ray sources with thermal-like spectra observed close to the centers of supernova remnants without any counterparts in radio and gamma wavebands. With blackbody temperatures of hundreds of eV and luminosities in the range $10^{33} - 10^{34}$ erg s$^{-1}$, they present no evidence of a pulsar magnetic field.
This scenario was recently revisited by \citet{Vigano-Pons2012} in the context of diffusion of the magnetic field post-hyperaccretion. Following these ideas, we perform hydrodynamic numerical simulations of the hypercritical phase for the Cas A scenario, focusing on the formation of a thin new crust built in such a phase. We found that the size of this crust is at the same height-scale of the neutrino-sphere where the neutrino cooling is operative. We show that a very necessary ingredient for forming an atmosphere in quasi-hydrostatic equilibrium is a sink of energy at the bottom of the envelope, which allows to deposit large amounts of material onto the neutron star's surface. The copious amount of neutrinos generated on the neutrino-sphere, at a very efficient rate, provides this support.\\
As a signature of this phase we calculate that the number of events expected on Hyper-Kamiokande detector is 3195.    This number of events going through the supernova remnant will oscillate, firstly in the thermal and magnetized plasma, secondly, due to different electron densities from two to four regions and finally, on their paths (vacuum) to Earth.    For the first region, we have used the neutrino effective potential derived in \citet{2014ApJ...787..140F} which is  a function of  temperature ($T$), chemical potential ($\mu$),  neutrino energy ($E_\nu$)  and magnetic field ($B$).   We have shown that for a neutrino test of energy 1, 5 , 10 and 20 MeV, and parameters considered of temperature and chemical potential in the range of 1 MeV $\leq T \leq$ 5 MeV,   $1\, {\rm eV} \leq \mu  \leq 4.3\times 10^2\, {\rm eV}$,  neutrinos oscillate resonantly, for two- and three-neutrino mixing.  In regions from two to four, we have also calculated each effective potential and then analyzed their oscillations through these regions.  Considering the values obtained in the simulations for Cas A, the neutrino flavor ratios expected on Earth were computed. Our analysis shows that deviations  from 1:1:1 are obtained  for neutrino energies of 5, 10, 15 and 20 MeV (see table 3).  Diverse flavor ratios of thermal neutrinos will give us information about the parameters involved in this hypercritical accretion episode. \\
RCW103, Pup A, and Kes 79 \citep{Kaspi2010} supernova remnants are good candidates to test the validity of the hypercritical phase and the subsequent submergence of the magnetic field, in the early history of these supernovae.\\
\section*{Acknowledgements}
We are grateful to DGTIC-UNAM and to IA-UNAM for allowing us to use their MIZTLI Cluster where all the simulations were performed. The software used in this work was in part developed by the DOE NNSA-ASC OASCR Flash Center at the University of Chicago. This work was supported in part by CONACyT grants CB-2009-1 No. 132400, CB-2008-1 No. 101958, project 128556-F and project 165584. Also, it was supported by PAPIIT project IN106212. Also we thank to John Beacom, Dany Page and William Lee for useful discussions. This work was supported by Luc Binette scholarship and the projects IG100414. C.G. Bernal is grateful to CAPES--Brazil for the postdoctoral fellowship received through the Science Without Borders program.


\begin{thebibliography}{}

\bibitem[\protect\citeauthoryear{{Abe} \& et al.}{{Abe} \&
  et~al.}{2011a}]{2011arXiv1109.3262A}
{Abe} K.,  et al. 2011a, ArXiv e-prints

\bibitem[\protect\citeauthoryear{{Abe} \& et al.}{{Abe} \&
  et~al.}{2011b}]{abe11a}
{Abe} K.,  et al. 2011b, Physical Review Letters, 107, 241801

\bibitem[\protect\citeauthoryear{{Aharmim} \& et al.}{{Aharmim} \&
  et~al.}{2011}]{aha11}
{Aharmim} B.,  et al. 2011, ArXiv e-prints

\bibitem[\protect\citeauthoryear{{Akhmedov}, {Johansson}, {Lindner}, {Ohlsson}
  \& {Schwetz}}{{Akhmedov} et~al.}{2004}]{akh04}
{Akhmedov} E.~K.,  {Johansson} R.,  {Lindner} M.,  {Ohlsson} T.,    {Schwetz}
  T.,  2004, Journal of High Energy Physics, 4, 78

\bibitem[\protect\citeauthoryear{{Araki} \& et al.}{{Araki} \&
  et~al.}{2005}]{ara05}
{Araki} T.,  et al. 2005, Physical Review Letters, 94, 081801

\bibitem[\protect\citeauthoryear{{Ashworth}
  Jr.}{{Ashworth}}{1980}]{1980JHA....11....1A}
{Ashworth} Jr. W.~B.,  1980, Journal for the History of Astronomy, 11, 1

\bibitem[\protect\citeauthoryear{{Bahcall}}{{Bahcall}}{1989}]{1989neas.book.....B}
{Bahcall} J.~N.,  1989, {Neutrino astrophysics}

\bibitem[\protect\citeauthoryear{{Bernal}, {Lee} \& {Page}}{{Bernal}
  et~al.}{2010}]{Bernal2010}
{Bernal} C.~G.,  {Lee} W.~H.,    {Page} D.,  2010, \rmxaa, 46, 309

\bibitem[\protect\citeauthoryear{{Bernal}, {Page} \& {Lee}}{{Bernal}
  et~al.}{2013}]{Bernal2013}
{Bernal} C.~G.,  {Page} D.,    {Lee} W.~H.,  2013, \apj, 770, 106

\bibitem[\protect\citeauthoryear{{Blondin}}{{Blondin}}{1986}]{Blondin1986}
{Blondin} J.~M.,  1986, \apj, 308, 755

\bibitem[\protect\citeauthoryear{{Chang} \& {Bildsten}}{{Chang} \&
  {Bildsten}}{2004}]{2004ApJ...605..830C}
{Chang} P.,  {Bildsten} L.,  2004, \apj, 605, 830

\bibitem[\protect\citeauthoryear{{Chevalier}}{{Chevalier}}{1989}]{Chevalier1989}
{Chevalier} R.~A.,  1989, \apj, 346, 847

\bibitem[\protect\citeauthoryear{{Chevalier}}{{Chevalier}}{2005}]{Chevalier2005}
{Chevalier} R.~A.,  2005, \apj, 619, 839

\bibitem[\protect\citeauthoryear{{Chevalier} \& {Soker}}{{Chevalier} \&
  {Soker}}{1989}]{che89}
{Chevalier} R.~A.,  {Soker} N.,  1989, \apj, 341, 867

\bibitem[\protect\citeauthoryear{{Dicus}}{{Dicus}}{1972}]{Dicus1972}
{Dicus} D.~A.,  1972, \prd, 6, 941

\bibitem[\protect\citeauthoryear{Forero, T\'ortola \& Valle}{Forero
  et~al.}{2012}]{PhysRevD.86.073012}
Forero D.~V.,  T\'ortola M.,    Valle J. W.~F.,  2012, Phys. Rev. D, 86, 073012

\bibitem[\protect\citeauthoryear{{Fraija}}{{Fraija}}{2014}]{2014ApJ...787..140F}
{Fraija} N.,  2014, \apj, 787, 140

\bibitem[\protect\citeauthoryear{{Fraija}, {Bernal} \&
  {Hidalgo-Gam{\'e}z}}{{Fraija} et~al.}{2014}]{2014MNRAS.442..239F}
{Fraija} N.,  {Bernal} C.~G.,    {Hidalgo-Gam{\'e}z} A.~M.,  2014b, \mnras,
  442, 239

\bibitem[\protect\citeauthoryear{{Fryxell}, {Olson}, {Ricker}, {Timmes},
  {Zingale}, {Lamb}, {MacNeice}, {Rosner}, {Truran} \& {Tufo}}{{Fryxell}
  et~al.}{2000}]{Fryxell2000}
{Fryxell} B.,  {Olson} K.,  {Ricker} P.,  {Timmes} F.~X.,  {Zingale} M.,
  {Lamb} D.~Q.,  {MacNeice} P.,  {Rosner} R.,  {Truran} J.~W.,    {Tufo} H.,
  2000, \apjs, 131, 273

\bibitem[\protect\citeauthoryear{{Geppert}, {Page} \& {Zannias}}{{Geppert}
  et~al.}{1999}]{Geppert1999}
{Geppert} U.,  {Page} D.,    {Zannias} T.,  1999, \aap, 345, 847

\bibitem[\protect\citeauthoryear{{Gonzalez-Garcia}}{{Gonzalez-Garcia}}{2011}]{gon11}
{Gonzalez-Garcia} M.~C.,  2011, Physics of Particles and Nuclei, 42, 577

\bibitem[\protect\citeauthoryear{{Gonzalez-Garcia} \&
  {Maltoni}}{{Gonzalez-Garcia} \& {Maltoni}}{2008}]{gon08}
{Gonzalez-Garcia} M.~C.,  {Maltoni} M.,  2008, \physrep, 460, 1

\bibitem[\protect\citeauthoryear{{Gonzalez-Garcia} \& {Nir}}{{Gonzalez-Garcia}
  \& {Nir}}{2003}]{gon03}
{Gonzalez-Garcia} M.~C.,  {Nir} Y.,  2003, Reviews of Modern Physics, 75, 345

\bibitem[\protect\citeauthoryear{{Ho} \& {Heinke}}{{Ho} \&
  {Heinke}}{2009}]{2009Natur.462...71H}
{Ho} W.~C.~G.,  {Heinke} C.~O.,  2009b, \nat, 462, 71

\bibitem[\protect\citeauthoryear{{Hyper-Kamiokande Working Group}, {:}, {Abe},
  {Aihara}, {Andreopoulos}, {Anghel}, {Ariga}, {Ariga}, {Asfandiyarov},
  {Askins} \& et al.}{{Hyper-Kamiokande Working Group}
  et~al.}{2014}]{2014arXiv1412.4673H}
{Hyper-Kamiokande Working Group} {:} {Abe} K.,  {Aihara} H.,  {Andreopoulos}
  C.,  {Anghel} I.,  {Ariga} A.,  {Ariga} T.,  {Asfandiyarov} R.,  {Askins} M.,
     et al. 2014, ArXiv e-prints

\bibitem[\protect\citeauthoryear{{Itoh}, {Hayashi}, {Nishikawa} \&
  {Kohyama}}{{Itoh} et~al.}{1996}]{Itoh1996}
{Itoh} N.,  {Hayashi} H.,  {Nishikawa} A.,    {Kohyama} Y.,  1996, \apjs, 102,
  411

\bibitem[\protect\citeauthoryear{{Kaspi}}{{Kaspi}}{2010}]{Kaspi2010}
{Kaspi} V.~M.,  2010, Proceedings of the National Academy of Science, 107, 7147

\bibitem[\protect\citeauthoryear{{Krause}, {Birkmann}, {Usuda}, {Hattori},
  {Goto}, {Rieke} \& {Misselt}}{{Krause} et~al.}{2008}]{Krause2008}
{Krause} O.,  {Birkmann} S.~M.,  {Usuda} T.,  {Hattori} T.,  {Goto} M.,
  {Rieke} G.~H.,    {Misselt} K.~A.,  2008, Science, 320, 1195

\bibitem[\protect\citeauthoryear{{Learned} \& {Pakvasa}}{{Learned} \&
  {Pakvasa}}{1995}]{lea95}
{Learned} J.~G.,  {Pakvasa} S.,  1995, Astroparticle Physics, 3, 267

\bibitem[\protect\citeauthoryear{{Mohapatra} \& {Pal}}{{Mohapatra} \&
  {Pal}}{2004}]{2004mnpa.book.....M}
{Mohapatra} R.~N.,  {Pal} P.~B.,  2004, {Massive neutrinos in physics and
  astrophysics}

\bibitem[\protect\citeauthoryear{{Muslimov} \& {Page}}{{Muslimov} \&
  {Page}}{1995}]{1995ApJ...440L..77M}
{Muslimov} A.,  {Page} D.,  1995, \apjl, 440, L77

\bibitem[\protect\citeauthoryear{{Pavlov}, {Sanwal} \& {Teter}}{{Pavlov}
  et~al.}{2004}]{Pavlov2004}
{Pavlov} G.~G.,  {Sanwal} D.,    {Teter} M.~A.,  2004, in {Camilo} F.,
  {Gaensler} B.~M.,  eds, Young Neutron Stars and Their Environments Vol.~218
  of IAU Symposium, {Central Compact Objects in Supernova Remnants}.
p.~239

\bibitem[\protect\citeauthoryear{{Reed}, {Hester}, {Fabian} \&
  {Winkler}}{{Reed} et~al.}{1995}]{1995ApJ...440..706R}
{Reed} J.~E.,  {Hester} J.~J.,  {Fabian} A.~C.,    {Winkler} P.~F.,  1995,
  \apj, 440, 706

\bibitem[\protect\citeauthoryear{{Sahu}, {Fraija} \& {Keum}}{{Sahu}
  et~al.}{2009a}]{2009PhRvD..80c3009S}
{Sahu} S.,  {Fraija} N.,    {Keum} Y.-Y.,  2009a, \prd, 80, 033009

\bibitem[\protect\citeauthoryear{{Sahu}, {Fraija} \& {Keum}}{{Sahu}
  et~al.}{2009b}]{2009JCAP...11..024S}
{Sahu} S.,  {Fraija} N.,    {Keum} Y.-Y.,  2009b, \jcap, 11, 24

\bibitem[\protect\citeauthoryear{{Shirai} \& {KamLAND Collaboration}}{{Shirai}
  \& {KamLAND Collaboration}}{2007}]{shi07}
{Shirai} J.,  {KamLAND Collaboration} 2007, Nuclear Physics B Proceedings
  Supplements, 168, 77

\bibitem[\protect\citeauthoryear{{the KamLAND Collaboration} \& {Mitsui}}{{The
  KamLAND Collaboration} \& {Mitsui}}{2011}]{mit11}
{The KamLAND Collaboration} {Mitsui} T.,  2011, Nuclear Physics B Proceedings
  Supplements, 221, 193

\bibitem[\protect\citeauthoryear{{Timmes} \& {Swesty}}{{Timmes} \&
  {Swesty}}{2000}]{2000ApJS..126..501T}
{Timmes} F.~X.,  {Swesty} F.~D.,  2000, \apjs, 126, 501

\bibitem[\protect\citeauthoryear{{Vigan{\`o}} \& {Pons}}{{Vigan{\`o}} \&
  {Pons}}{2012}]{Vigano-Pons2012}
{Vigan{\`o}} D.,  {Pons} J.~A.,  2012, \mnras, 425, 2487

\end{thebibliography}

%
\clearpage
\clearpage
\end{document}